\documentclass[12pt,a4paper]{article}\usepackage[]{graphicx}\usepackage[]{xcolor}
\makeatletter
\def\maxwidth{ %
  \ifdim\Gin@nat@width>\linewidth
    \linewidth
  \else
    \Gin@nat@width
  \fi
}
\makeatother

\definecolor{fgcolor}{rgb}{0.345, 0.345, 0.345}

\usepackage{framed}
\makeatletter
 {\par\unskip\endMakeFramed%
 \at@end@of@kframe}
\makeatother

\definecolor{shadecolor}{rgb}{.97, .97, .97}
\definecolor{messagecolor}{rgb}{0, 0, 0}
\definecolor{warningcolor}{rgb}{1, 0, 1}
\definecolor{errorcolor}{rgb}{1, 0, 0}

\usepackage{alltt}

\usepackage{amsmath, amsfonts, amssymb, amsthm}
\usepackage[english]{babel}

\usepackage{url}
\usepackage{fullpage}
\usepackage{hyperref}
\usepackage{color}
\usepackage{orcidlink}

\usepackage[authoryear,round,longnamesfirst]{natbib}
\bibpunct{(}{)}{;}{a}{}{,}

\makeatletter
\newcommand\code{\bgroup\@makeother\_\@makeother\~\@makeother\$\@codex}
\def\@codex#1{{\normalfont\ttfamily\hyphenchar\font=-1 #1}\egroup}
\makeatother

\newcommand{\pkg}[1]{{\fontseries{b}\selectfont #1}}

\newcommand{\m}{\mathbf}   
\newcommand{\bs}{\boldsymbol}
\newcommand{\Lb}[1]{#1_{\texttt{L}}}
\newcommand{\Ub}[1]{#1_{\texttt{U}}}
\newcommand{\Prof}[1]{#1_{\texttt{prof}}}
\newcommand{\New}[1]{#1_{\texttt{new}}}
\newcommand{\Prolang}[1]{\textsf{#1}}
\definecolor{InputColor}{rgb}{0.600,0.060,0.360} 
\definecolor{OutputColor}{rgb}{0.133,0.543,0.133}
\definecolor{Gray}{rgb}{0.5,0.5,0.5}
  {\medskip \par}
\title{\bf
  Profile Likelihood via Optimisation and Differential Equations
}

\author{Yves Deville$^\star$}
\date{\today}
\IfFileExists{upquote.sty}{\usepackage{upquote}}{}
\begin{document}

\maketitle{}

\begin{center}\centering \footnotesize
  $^\star$Statistical Consultant, Chambéry FR.\\
  Email:~\texttt{deville.yves@alpestat.com}.
  \orcidlink{https:/orcid.org/0000-0002-1233-488X}
\end{center}

\newtheorem{theo}{Proposition}

\newtheoremstyle{remark}
{0.4em}
{0.4em}
{\small}
{0em}
{\bf}
{.}
{ }
{}
\renewcommand{\qedsymbol}{$\blackksquare$}
\theoremstyle{remark}
\newtheorem{rk}{Remark}

\setcounter{page}{1}

\setkeys{Gin}{width=7.4cm}

\abstract{

  Profile likelihood provides a general framework to infer on a scalar parameter
  of a statistical model. A confidence interval is obtained by numerically
  finding the two abscissas where the profile log-likelihood curve intersects an
  horizontal line. An alternative derivation for this interval can be obtained
  by solving a constrained optimisation problem which can broadly be described
  as: maximise or minimise the parameter of interest under the constraint that
  the log-likelihood is high enough. This formulation allows nice geometrical
  interpretations; It can be used to infer on a parameter or on a known scalar
  function of the parameter, such as a quantile.  Widely available routines for
  constrained optimisation can be used for this task, as well as Markov Chain
  Monte Carlo samplers. When the interest is on a smooth function depending on
  an extra continuous variable, the constrained optimisation framework can be
  used to derive Ordinary Differential Equation (ODE) for the confidence
  limits. This is illustrated with the return levels of Extreme Value models
  based on the Generalised Extreme Value distribution. Moreover the same
  ODE-based technique applies as well to the derivation of profile likelihood
  contours for couples of parameters. The initial value of the ODE used in the
  determination of the interval or the contour can itself be obtained by another
  auxiliary ODE with known initial value obtained by using the confidence
  level as the extra continuous variable.

}


\section{Problem}
We consider here a regular parametric statistical model depending on a
vector~$\bs{\theta}$ of $p$ parameters, and focus on the inference on a specific
scalar component of the parameter using a vector of observations $\m{y}$,
possibly depending on covariates. Without loss of generality we focus on the
first component denoted by $\psi$, while $\bs{\lambda}$ will be the vector of
the remaining $p-1$ components, so
$\bs{\theta} = [\psi,\,\bs{\lambda}^\top]^\top$.  The log-likelihood function
$\ell(\bs{\theta};\,\m{y})$ will be denoted as $\ell(\bs{\theta})$ or
$\ell(\psi,\,\bs{\lambda})$ and will be assumed to be differentiable twice at
least.  We will assume that the log-likelihood is ``well-shaped'': it has a
unique maximum $\hat{\bs{\theta}}$ with components $\hat{\psi}$ and
$\hat{\bs{\lambda}}$, and moreover the profile log-likelihood denoted by
$\Prof{\ell}(\psi)$ will be assumed to be increasing in $\psi$ for
$\psi \leqslant \hat{\psi}$, and to be decreasing for
$\psi \geqslant \hat{\psi}$. We denote somewhat abusively by $\psi(\bs{\theta})$
and $\bs{\lambda}(\bs{\theta})$ the sub-vectors of $\bs{\theta}$ with respective
length $1$ and $p-1$.

\begin{figure}
  \centering
  \includegraphics[width=0.7\textwidth]{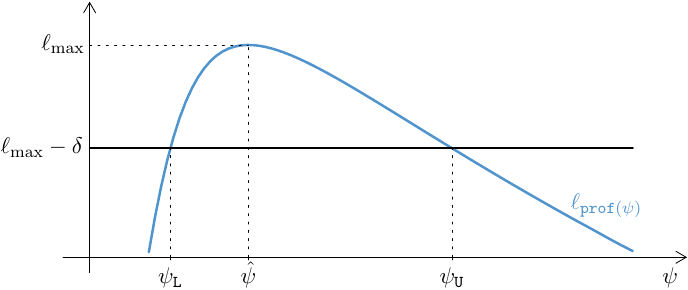}
  \caption{\label{FigProfLik} \sf \footnotesize Profile
    likelihood interval in the
    ``well-shaped'' case. }
\end{figure}

Using our assumption that the log-likelihood is well-shaped, we can denote
by~$\Ub{\psi}$ the upper bound of the likelihood-profile confidence interval
with level $1 - \alpha$. The value~$\Ub{\psi} > \hat{\psi}$ is a solution of the
equation
\begin{equation}
  \label{eq:ellDelta}
  \Prof{\ell}(\psi) = \ell_{\max} - \delta
\end{equation}
where $\ell_{\max} = \ell(\hat{\bs{\theta}})$ is the maximal log-likelihood, and
where $\delta>0$ relates to the quantile of the $\chi^2$ distribution with one
degree of freedom according to $\delta = q_{\chi^2(1)}(1 - \alpha) /
2$. Similarly the lower bound $\Lb{\psi}$ solves the
equation~(\ref{eq:ellDelta}) with $\Lb{\psi} < \hat{\psi}$, see
Figure~\ref{FigProfLik}. The specific choice of~$\delta$ grants that the
confidence interval \textit{asymptically} reaches the prescribed confidence
level.

The practical determination of profile likelihood intervals can rely on
what we may call a \textit{naive} method: the profile likelihood
function $\Prof{\ell}(\psi)$ being implemented by using a numerical
$(d-1)$-dimensional optimisation, one can solve~(\ref{eq:ellDelta}) by
using a zero-finding method. \cite{Venzon_Profile} instead designed an
algorithm to find the bounds of the confidence interval by solving a
nonlinear equation in $\bs{\theta}$, hence without computing the
profile likelihood as such.  Quite recently, some authors have
designed implementations of the profile likelihood inference by
relying on a formulation as a constrained optimisation problem in
which the objective is the parameter of interest, realising that the
non-linear equation used by Venzon and Moolgavkar encodes the
Karush-Kuhn-Tucker conditions for this optimisation problem. While
this potentially makes the use of the profile likelihood method nearly
as simple as that of the cheap delta method, practical problems arise
related to the optimisation. \cite{FisherLewis_Profile} have designed
a dedicated optimisation algorithm called RVM for \textit{Revised
  Venzon-Moolgavkar}. They have studied a variety of strategies
regarding the numerical optimisation and have found that the RVM
outperforms the original Venzon-Moolgavkar method and the other methods
on a large benchmark of problems. \cite{BorisovMetelkin_Profile} also
describe the use of optimisation in a software computing profile
likelihood confidence intervals on a parameter or on a function of the
parameter vector. They discuss the question of practical
identifiability in relation with a specific form of model as used in
biology.

Our motivation is for models related to the Extreme Value (EV) theory, such as
models for block maxima, with margins following the Generalised Extreme-Value
(GEV) distribution. For these models, the confidence intervals based on profile
likelihood are known to have a much better coverage rate than those based on the
delta method or on the bootstrap. We will focus on the use of the \Prolang{R}
language~\citep{R} for the implementation. In practice, the profile likelihood
intervals are often computed via the naive method as is the case for the famous
R packages available on the Comprehensive R Archive Network (CRAN): \pkg{ismev}
\citep{pack_ismev}, \pkg{evd} \citep{pack_evd}, \pkg{extRemes}
\citep{pack_extRemes}, \pkg{mev} \citep{pack_mev} and others. Beside the model
parameters, one often has to infer on the return levels corresponding to large
return periods such as $100$ or even $1000$ years, and some tedious
re-parameterisation can then be required. For EV models, a Bayesian predictive
approach is nowadays often preferred to the frequentist approach since it
provides more easily interpreted results and eventually avoids the use of return
levels. However the likelihood inference has the advantage of being independent
of the specific parameterisation used, which is not the case for the Bayesian
inference, even when a non-informative prior is used. To our best knowledge,
using constrained optimisation methods for the inference on EV models has not
yet been reported. We have been experimenting on this for several years and came
up to some methods relying on the use of Ordinary Differential Equation (ODE) as
described in sections~\ref{SecExtraODE} and~\ref{SecODEContours}

This paper is organised as follows.  In section~\ref{SecOptim} we
recall the constrained optimisation formulation. In
section~\ref{SecFunction} we focus on the inference on a function of
the parameter, illustrate it on examples and discuss some practical
issues.  In Section~\ref{SecExtraODE} we consider a function depending
on an extra ``time'' variable and show how an ODE then naturally
arises. Section~\ref{SecODEContours} describes some methods closely
related to the geometry of the profile likelihood contours or of the
likelihood contours: ODEs can be used in relation with a suitable time
variable with geometrical interpretation.  Finally,
Section~\ref{SecConcl} and discusses on limitations and possible
extensions. 

\section{Optimisation problems}
\label{SecOptim}
  
The profile log-likelihood function $\Prof{\ell}(\psi)$ is defined by
\begin{equation}
  \Prof{\ell}(\psi) := \ell\{\psi, \, \hat{\bs{\lambda}}(\psi) \}
\end{equation}
where $\hat{\bs{\lambda}}(\psi)$ maximises the function
$\bs{\lambda} \mapsto \ell(\psi,\,\bs{\lambda})$.
So for a given $\psi$, the
vector $[\psi,\,\hat{\bs{\lambda}}(\psi)^\top]^\top$ is the solution of an
optimisation problem with equality constraint
\begin{equation}
  \label{eq:OptimConstrPsi}
  \begin{aligned}
    & \underset{\bs{\theta}}{\textrm{maximise}} & & \ell(\bs{\theta}) \\ 
    & \textrm{subject to} & &  \psi(\bs{\theta}) = \psi. 
  \end{aligned}
\end{equation}
With the assumptions above, we can regard the upper bound $\Ub{\psi}$ of
the confidence interval on $\psi$ as the solution of
\begin{equation}
  \begin{aligned}
  & \underset{\psi}{\text{maximise}} & & \psi \\ 
  &  \textrm{subject to} & & \Prof{\ell}(\psi) \geqslant 
  \ell_{\max} - \delta.
  \end{aligned}
\end{equation}
Indeed, $\Ub{\psi}$ is the largest $\psi$ for which the inequality
$\Prof{\ell}(\psi) \geqslant \ell_{\max} - \delta$ holds.  The constraint
function of this problem is itself defined as the solution of an optimisation
problem, which limits its practical use. Yet we can instead solve the following
optimisation problem which uses the log-likelihood function as constraint
function.

\begin{theo}{}
  \label{theo:Optim}
  Let $\psi^\star$ be defined as $\psi(\bs{\theta}^\star)$ where
  $\bs{\theta}^\star$ with components $\psi^\star$ and
  $\bs{\lambda}^\star$ is the solution of the constrained optimisation
  problem
  
  \begin{equation}
    \label{eq:PbStar}
    \begin{aligned}
      & \underset{\bs{\theta}}{\text{maximise}} & & \psi(\bs{\theta}) \\ 
      &  \textrm{subject to} & & \ell(\bs{\theta}) 
      \geqslant \ell_{\max} - \delta,
    \end{aligned}
  \end{equation} 
  where $\delta := q_{\chi^2(1)}(1 - \alpha) / 2$.  Then $\psi^\star$ is
  nothing but the upper end-point $\Ub{\psi}$ of the $100(1 - \alpha)\%$
  confidence interval obtained by profile likelihood.
  Similarly, if $\bs{\theta}_\star$ with components $\psi_\star$ and
  $\bs{\lambda}_\star$ is the solution of the \emph{minimisation} of
  $\psi(\bs{\theta})$ under the same constraint, then $\psi_\star$ is
  the lower bound $\Lb{\psi}$ of the profile likelihood
  confidence interval.
  
\end{theo}

A proof for this simple result is given in Appendix~\ref{Proof}.  An appealing
geometrical derivation for the case $p =2$ is as follows.  The profile
log-likelihood for $\psi$ is the profile of a surface in the usual meaning,
namely the log-likelihood surface. This profile is the curve that would be seen
by an observer located at infinity in the direction of the $\lambda$-axis, at
$\lambda = -\infty$. By cutting the surface at the level $\ell_{\max} - \delta$
we define a contour projecting on the $\psi\lambda$-plane into the contour of
the feasible set, say $\mathcal{R}(\delta)$, of
problem~(\ref{eq:PbStar}). Obviously, our observer can locate $\Lb{\psi}$ and
$\Ub{\psi}$ as the value of $\psi$ for the leftmost and the rightmost point that
can be seen in $\mathcal{R}$. These are identical to the values where the
profile is seen to cut the horizontal line corresponding to the altitude
$\ell_{\max} - \delta$.

\begin{rk}{}
  \label{rk:Concentr}
  Suppose that the ML estimate is available in closed form for one component of
  $\bs{\lambda}$, for instance $\lambda_1$. Then in the optimisation
  problem~(\ref{eq:PbStar}) we can replace the vector $\bs{\theta}$ by the
  vector $\bs{\theta}_{\lambda_1^{\text{c}}}$ with length $p-1$ obtained by
  discarding~$\lambda_1$ from $\bs{\theta}$ and replace the log-likelihood
  $\ell(\bs{\theta})$ in the constraint by the profile likelihood function
  $\ell_{\lambda_1^{\text{c}}}(\bs{\theta}_{\lambda_1^{\text{c}}})$ obtained
  from $\ell(\bs{\theta})$ by replacing $\lambda_1$ by its ML
  estimate~$\hat{\lambda}_1(\bs{\theta}_{\lambda_1^{\text{c}}})$. The main point
  here is that the objective $\psi(\bs{\theta}) = \psi$ does not depend on
  $\lambda_1$. An example will be given in section~\ref{SecLinReg} below.

\end{rk}
\begin{rk}{}
  \label{rk:IneqOrEq}
  In the constraint of~(\ref{eq:PbStar}) the equality 
  $\ell(\bs{\theta}) = \ell_{\max} - \delta$ can be used in place
  of the inequality.

\end{rk}

\section{Function of the parameter vector}
\label{SecFunction}

\subsection{Function value vs. parameter}
\label{SecpConf}
In many situations we want to infer on a scalar smooth function of the parameter
vector, say~$\eta(\bs{\theta})$. A typical example is when $\eta$ is the
expectation of a response for some ``new'' value of a vector of covariates. The
profile likelihood method can then be used.

An usual approach is to consider $\eta(\bs{\theta)}$ as a scalar parameter using
a suitable re-parameterisation of the model that discards one the original
parameters and replaces it by $\eta$. Mathematical conditions for the existence
of such a re-parameterisation seem quite complex and will not be discussed
here. Yet if this re-parameterisation is admissible, we are exactly in the
situation of the previous sections with $\psi(\bs{\theta})$ replaced by
$\eta(\bs{\theta})$.

There are some difficulties in the previous approach. Needless to say, a careful
analysis is required to choose the discarded parameter.  From a software
engineering point of view, this can be tedious because models are often defined
using formulas, and these are not easily re-parameterised. The
re-parameterisation can also have a negative impact on the numerical
conditioning e.g., when the new parameter turns out to be strongly correlated to
those kept in the model.  Moreover, the interest can be both on the parameters
and on a function of these, or on several functions of the parameters: Using a
regression model, several predictions are often considered. With several
functions, we have to use as many re-parameterisations. By contrast, it is much
simpler to rely on Proposition~\ref{theo:Optim} and use an optimisation program
with the objective function being~$\eta(\bs{\theta})$ instead of
$\psi(\bs{\theta})$. A constrained optimisation algorithm should perform
automatically a good local re-parameterisation based on the value of the
iterate, and this is likely to work better than the global re-parametrisation
considered above.

An approximate confidence region $\mathcal{R}$ on the the
full parameter vector $\bs{\theta}$ is obtained as 
\begin{equation}
  \label{eq:Rdelta}
  \mathcal{R}(\delta) := \left\{ \bs{\theta}: \,  \ell(\bs{\theta}) 
    \geqslant \ell_{\max} - \delta \right \}
\end{equation}
with $\delta := q_{\chi^2(p)}(1 - \alpha) / 2$. Quite obviously, a so-called
\textit{conservative} confidence interval on $\eta$ can be obtained by
minimising and maximising $\eta(\bs{\theta})$ for $\bs{\theta} \in \mathcal{R}$
to get the confidence limits $\Lb{\eta}$ and $\Ub{\eta}$.  A striking point is
that Prop.~\ref{theo:Optim} tells that we can replace in the definition of
$\delta$ the quantile $q_{\chi^2(p)}$ of the $\chi^2$ distribution with $p$
degree of freedom by the equivalent quantile $q_{\chi^2(1)}$ for \textit{one}
degree of freedom (d.f.), leading to a \emph{much smaller} confidence interval.
An essential point is that $\eta(\bs{\theta})$ must be such that it can be used
as a parameter for the model; This is a condition of identiability.

In the Bayesian framework, the inference on a function of the parameter, be it
scalar or vector, is straightforward. Indeed if Monte Carlo Markov Chain (MCMC)
iterates $\bs{\theta}^{[k]}$ are available, the values $\eta(\bs{\theta}^{[k]})$
can be computed to provide a sample of the posterior distribution for
$\eta(\bs{\theta})$. Credible intervals on $\eta(\bs{\theta})$ can then be
given. However, a prior for $\bs{\theta}$ is involved in this process. By using
a non-informative ``flat'' prior with $p(\bs{\theta}) \propto 1$ (most often
improper), the posterior density will be proportional to the likelihood. Yet the
prior for $\bs{\theta}$ induces a prior for $\eta(\bs{\theta})$ which may be
informative, especially when the available observations carry little information
about~$\eta(\bs{\theta})$. An example is provided by a regression model when
$\eta(\bs{\theta})$ is a new response at a design point far away from the design
points used in the fit. As detailed in Section~\ref{SecMCMC}, MCMC iterates can
be used for a frequentist likelihood-based inference no longer
depending on a prior choice.

\subsection{Lagrangian formulation}
Although the optimisation problem will in practice be most of time solved by
using an optimisation software, the Lagrangian $\mathcal{L}(\bs{\theta},\,\nu)$
provides insights about the problem. It is defined by
\begin{equation}
  \label{eq:Lagrangian}
  \mathcal{L}(\bs{\theta},\,\nu) := \eta(\theta) - 
  \nu \left\{ \rule{0pt}{1em} \ell(\bs {\theta}) - \ell_{\textrm{max}} + \delta\right\}
\end{equation}
where $\nu$ is a Lagrange multiplier. The first-order condition
a.k.a. as \textit{Karush-Kuhn-Tucker (KKT) conditions} is obtained by
zeroing the derivatives of $\mathcal{L}(\bs{\theta},\,\nu)$ w.r.t. to
$\bs{\theta}$ and to $\nu$.  It writes as
\begin{subequations}
  \begin{align}
    \label{eq:Lagrangea}
    \frac{\partial \eta(\bs{\theta})}{\partial \bs{\theta}} &= \nu \, 
    \frac{\partial \ell(\bs{\theta})}{\partial \bs{\theta}}  \\ 
    \label{eq:Lagrangeb}
    \ell(\bs {\theta}) &=  \left\{\ell_{\textrm{max}} - \delta \right\}.
    \rule{0pt}{1.1em}
  \end{align}
\end{subequations}
The second equation hopefully defines a $(d-1)$-dimensional sub-manifold of
$\mathbb{R}^d$, namely: a contour of the log-likelihood function. This is the
boundary $\partial \mathcal{R}$ of the confidence region~(\ref{eq:Rdelta})
defined above. The first condition tells that the gradient vector of the
objective function~$\eta$ is orthogonal to the tangent hyperplane of the
manifold $\partial\mathcal{R}$. As a general rule, two solutions
$\Ub{\bs{\theta}}$ and $\Lb{\bs{\theta}}$ exist for the set of two equations,
corresponding to the upper and lower bounds of the confidence interval.

In the special case where $\eta(\bs{\theta})$ is the first element~$\psi$
in~$\bs{\theta}$, the first of the $p$~scalar equations in (\ref{eq:Lagrangea})
gives~$\nu$ as the inverse of $\partial_{\psi} \ell$, and the $p-1$ remaining
equations can be written as $\partial_{\bs{\lambda}} \ell = \m{0}$. So in this
case the equations (\ref{eq:Lagrangea}) and (\ref{eq:Lagrangeb}) are nothing but
the formulation of \cite{Venzon_Profile}.

Assuming that the derivatives of $\eta(\bs{\theta})$ and that of
$\ell(\bs{\theta})$ can be used in closed form, one can in some cases eliminate
$\nu$ from the~$p + 1$ coordinate equations of~(\ref{eq:Lagrangea}) and
(\ref{eq:Lagrangeb}). For instance, assuming that the partial derivative
of~$\partial_{\theta_1} \eta$ w.r.t the first component $\theta_1$ does nor
vanish, the first coordinate equation can be used to divide both sides of the
$p-1$ remaining scalar equations in~(\ref{eq:Lagrangea}), thus eliminating $\nu$
from these. Then, with~(\ref{eq:Lagrangeb}) we get a set of $p$ scalar equations
for the vector~$\bs{\theta}$ --\,see section~\ref{GEVt} later for an example.

\subsection{Example: prediction in linear regression}
\label{SecLinReg}
Although of limited practical interest, a simple illustration comes from the
well-known linear regression $y = \m{x}^\top \bs{\theta} + \varepsilon$ where
$\m{x}$ is the vector of $p$ covariates and $\varepsilon$ if a normal
disturbance with mean zero and variance $\sigma^2$. We will temporarily consider
$\sigma^2$ as known.  We further assume that $n \geqslant p$ observations are
available for the estimation, and that the design matrix $\m{X}$ with dimension
$n \times p$ has full column rank.

Taking $\eta$ as being the prediction mean for some new vector of covariates
$\New{\m{x}}$, namely $\eta := \New{\m{x}}^\top \bs{\theta}$, we can find a
confidence region by minimising and maximising $\eta(\bs{\theta})$ for
$\bs{\theta} \in \mathcal{R}(\delta)$, where the region $\mathcal{R}(\delta)$ of
the parameter space as defined by~(\ref{eq:Rdelta}) is here an ellipsoid with
centre $\hat{\bs{\theta}}$. The geometrical interpretation is easy: in the
parameter space, the levels of the prediction correspond to the family of the
hyperplanes which are orthogonal to the vector~$\New{\m{x}}$. We are looking for
the ``most extremes'' hyperplanes among those intersecting the ellipsoid
$\mathcal{R}(\delta)$. It is geometrically clear that two such hyperplanes
exist, both being tangent to the boundary $\partial \mathcal{R}$.  Using simple
algebra, the Lagrangian~(\ref{eq:Lagrangian}) is found to have the form
$$
 \mathcal{L}(\bs{\theta},\,\nu)  = \New{\m{x}}^\top \bs{\theta}  - \nu \left\{
   -\frac{1}{2} \,
 (\bs{\theta} - \hat{\bs{\theta}})^\top  \m{C}^{-1}
 ( \bs{\theta} - \hat{\bs{\theta}}) + \delta  \right\}
$$
where $\hat{\bs{\theta}}$ is the estimate of $\bs{\theta}$ and
$\m{C}:= \sigma^2 (\m{X} \m{X}^\top)^{-1}$ is its covariance matrix. The
relation~(\ref{eq:Lagrangea}) now writes as
$\bs{\theta} = \hat{\bs{\theta}} - \nu^{-1} \m{C}\,\New{\m{x}}$ and then the
constraint~(\ref{eq:Lagrangeb}) gives two values $\Ub{\nu}$ and $\Lb{\nu}$ for
the Lagrange multiplier: $\nu = \pm s_\mu(\New{\m{x}}) / \sqrt{2\delta}$, where
$s_{\mu}(\m{x}) := \{\m{x}^\top\m{C}\,\m{x}\}^{1/2} = \sigma \{\m{x}^\top
(\m{X}^\top\m{X})^{-1}\m{x}\}^{1/2}$ is the usual standard error for the
regression mean. Corresponding to the two vectors $\Ub{\bs{\theta}}$ and
$\Lb{\bs{\theta}}$, the two confidence limits $\Ub{\eta}$ and $\Lb{\eta}$ for
the predicted value are given by
$$
   \eta = \New{\m{x}}^\top \, \hat{\bs{\theta}}  \pm 
   s_{\mu}(\New{\m{x}}) \sqrt{2\delta}.
$$
Since $q_{\chi^2(1)}(1 - \alpha) = q_{Z}^2(1 - \alpha/2)$ when $Z$ follows a
standard normal distribution, we get the classical result of linear regression
textbooks. Note that this corresponds to using a $\chi^2$ distribution with
\textit{one} d.f., while using $p$ d.fs as considered in Section~\ref{SecpConf}
would lead to an over-conservative interval.

If $\sigma^2$ is considered as unknown, then the confidence interval will take
the same form, but with $\sigma^2$ replaced by its ML estimate
$\widehat{\sigma}^2$ i.e., by the mean of the squared residuals, so using the
denominator~$n$ rather than~$n-p$ in the formula for the empirical variance. We
are indeed in the situation described in Remark~\ref{rk:Concentr}: the
prediction~$\eta$ does not depend on the variance $\sigma^2$ but only on
the regression coefficients. Note that the confidence interval will only
asymptotically reach the target coverage rate because it uses the quantile of
the thin-tailed normal distribution, while a quantile of the thick-tailed
Student distribution~$t(n -p)$ with $n-p$ d.f. would be required to get an exact
interval.

\subsection{Return level for GEV}
\label{SecGEVRL}
\subsubsection*{The GEV distribution}

Recall that the GEV distribution~\citep{Coles_ISMEV} depends on a vector
$\bs{\theta} = [\mu,\,\sigma,\,\xi]^\top$ of three parameters called: the
\textit{location}~$\mu$, the \textit{scale} $\sigma >0$ and the \textit{shape}
$\xi$.  The distribution function is given by
$$
F(y,\,\bs{\theta}) = \begin{cases}
  \exp\{ - [1 + \xi (y - \mu) / \sigma ]_+^{-1/\xi} \} & \xi \neq 0,\\
  \exp\left\{-\exp( - [y - \mu] / \sigma) \right\}& \xi = 0 \rule{0pt}{1em}
\end{cases}
$$
where $z_+ := \max\{0,\,z\}$ for $z$ real.
For $\xi < 0$ the distribution has a finite upper end-point
$\omega = \mu - \sigma /\xi$. For $\xi > 0$ the distribution has a
long-tail. For $\xi = 0$ we get the thin-tailed Gumbel distribution.

This distribution is often used to describe so-called \textit{block maxima}
random variables such as annual maxima of temperature, rainfall, ... The MLE
$\hat{\bs{\theta}}$ of $\bs{\theta}$ is usually found by using a standard
numerical maximisation of the log-likelihood function $\ell(\bs{\theta})$.  Yet
it is worth noting that the constraint $\xi \geqslant - 1$ should be imposed
because we always have $\ell_{\max} = \infty$ when $\xi < -1$. Also, the
regularity conditions required for ML inference only hold when $\xi > -1/2$.

\begin{figure}
  \centering
  \begin{tabular}{cc}
    \includegraphics[width=8.6cm]{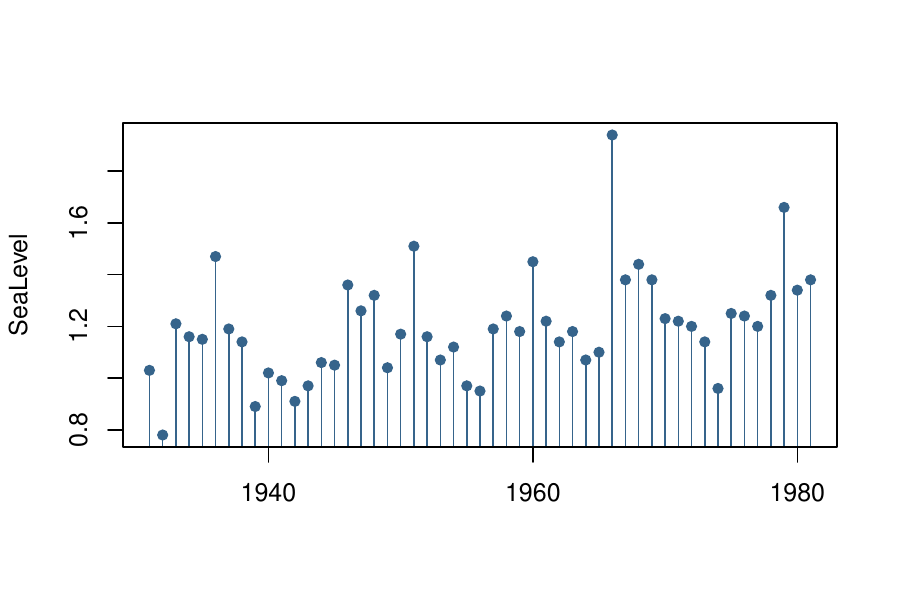}
    & \includegraphics[width=6cm]{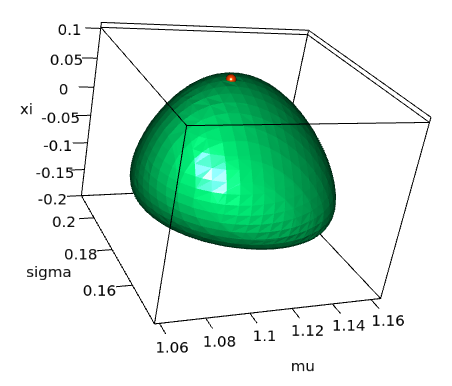}
  \end{tabular}                                                          
  \caption{\label{FigGEV} \sf \small Venice example. Left: The
    observations. Right: The contour $\partial \mathcal{R}$ in the
    parameter space with coordinates $\mu$, $\sigma$ and $\xi$; The
    little ball in orange on top of the contour surface shows the
    parameter $\bs{\theta}$ for which the shape $\xi$ is maximal. The
    corresponding coordinate $\xi$ is the upper bound
    $\xi_{\texttt{U}}$ of the confidence interval for $\xi$.  }
\end{figure}

For illustration we will use annual maxima $y_i$ for the sea level in
Venice \cite{Coles_ISMEV}. The maxima (in metres) are extracted from
the \code{venice} data shipped with the R package \pkg{ismev} and
rescaled. The simple model
$y_i \sim_{\text{i.i.d.}} \text{GEV}(\mu,\,\sigma,\,\xi)$ will be used
for illustration under the name \textit{Venice example}.  The $95\%$
interval on the shape parameter $\xi$ as found by constrained
optimisation is $[-0.197,\, 0.098]$. The right part of
Figure~\ref{FigGEV} shows the boundary $\partial \mathcal{R}$ of the
region $\mathcal{R}(\delta)$ of~(\ref{eq:Rdelta}) in the
three-dimensional parameter $\mu\sigma\xi$-space with
$\delta := q_{\chi^2(1)}(1 - \alpha) /2$. Also shown is the point
$\bs{\theta}$ for which $\xi$ is maximal, corresponding to the upper
confidence limit. All 3D plots shown in this article were created by
using the R package \pkg{rgl}~\citep{pack_rgl}.

\subsubsection*{Inferring on a return level}
For a given real number $T > 1$ called the \textit{return period}, we will
consider the corresponding \textit{return level} as denoted
by~$\eta(T; \,\bs{\theta})$ or simply $\eta(T)$, and defined by
\begin{equation}
  \label{eq:RL}
  \eta(T) = \mu + \sigma \, \eta_0(T), \qquad
  \text{with } \eta_0(T):= \begin{cases}
    \{ T ^{\xi} - 1\} / \xi& \xi \neq 0, \\
    \log T   & \xi = 0.
  \end{cases}
\end{equation}
This is the level which is on average exceeded once in a period with duration
$T$ years. 

\begin{rk}
  This definition of the return level is derived from a continuous-time
  framework rather than from a framework of blocks with a specific duration; It
  departs slightly from the usual definition of the return level as being the
  quantile $q_{\texttt{GEV}}(p)$ corresponding to the probability
  $p := 1 - 1 /T$ see~\cite{Coles_ISMEV}. However the two formulas are the same
  up to the approximation $- \log[- \log p] \approx -\log (1 - p) $ which holds
  for $p$ close to~$1$ hence for large~$T$.
\end{rk}

\begin{figure}
  \centering
  \includegraphics[width=0.86\textwidth]{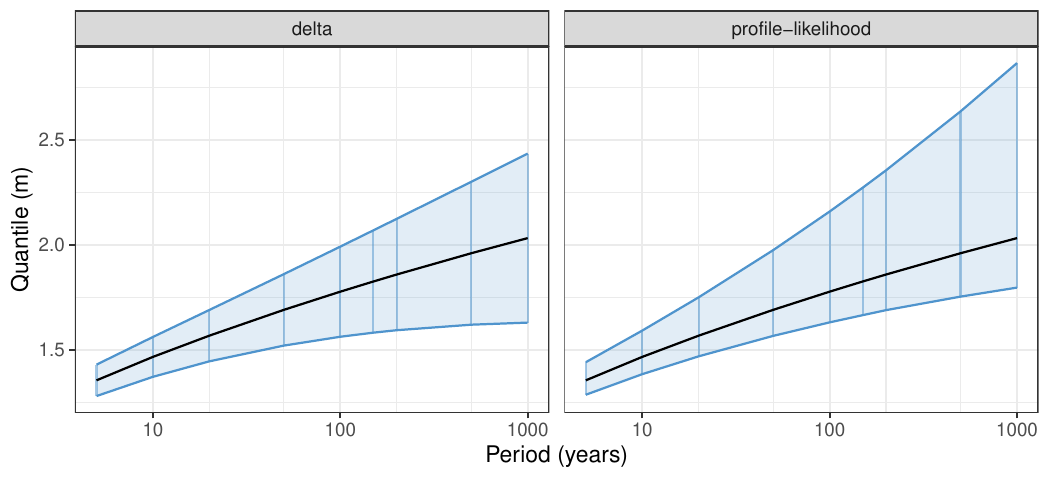}
  \caption{\label{FigVeniceRLPlot} \sf \footnotesize Return level plot for the
    Venice example using the cheap delta method (left) and profile likelihood
    (right).  The $95\%$ confidence bounds $\eta_{\texttt{L}}(T)$ and
    $\eta_{\texttt{U}}(T)$ are computed for the return periods~$T_i$ shown by a
    vertical segment. The curve in black shows the return level $\eta(T)$
    corresponding to the ML estimate $\hat{\bs{\theta}}$.}
\end{figure}

For a given return period~$T$, it is straightforward to re-parameterise the
model by replacing the location parameter~$\mu$ by~$\eta(T)$.  Both the
transformation $[\mu,\,\sigma,\,\xi]^\top \mapsto [\eta(T),\,\sigma,\,\xi]^\top$
and its reciprocal are easily coped with. However, we usually need to consider
several return periods. For example, a popular tool is the so-called
\textit{return level plot} as shown on Figure~\ref{FigVeniceRLPlot} for the
Venice example. The confidence bounds $\eta_{\texttt{L}}(T)$ and
$\eta_{\texttt{U}}(T)$ are actually obtained by interpolating the bounds
computed at a dozen of return periods~$T_i$. This was done both for the cheap
delta method and for the profile likelihood method using optimisation. Needless
to say, the two inference methods produce quite different results.

\begin{rk}
  For models specifying independent GEV observations, the gradient and Hessian
  of the log-likelihood $\ell(\bs{\theta})$ can be obtained in closed
  form. However, much care must be taken for~$\xi \approx 0$ because the exact
  derivatives are then quite difficult to evaluate, especially at the second
  order. The gradient of the objective can usually be obtained in closed form,
  as is obviously the case when the objective is one of the parameters. For the
  inference on the GEV return period, the gradient of the quantile is difficult
  to evaluate for $\xi \approx 0$. We used here Taylor approximations as
  implemented in the R package \pkg{nieve}~\citep{pack_nieve}.
\end{rk}

\subsubsection*{Extreme Value Regression}

The GEV distribution is often used in the so-called \textit{EV regression}
framework where a response~$y$ is related to a vector of covariates~$\m{x}$
$$
y \, \vert\, \m{x} \sim\text{GEV}\{\mu(\m{x}), \, \sigma(\m{x}),\, \xi(\m{x})\},
$$
where the functions $\mu(\m{x})$, $\sigma(\m{x})$ and $\xi(\m{x})$ can
be specified in parametric form.  As a fairly simple example to
investigate climate change from block maxima considered as being
independent, a scalar time~$x$ can be used as covariate for the
location parameter as in
\begin{equation}
  \label{eq:LinEV}
  \mu(x) = \theta_{0}^\mu + \theta_1^\mu x
\end{equation}
so the vector $\bs{\theta}$ of model parameters involves a
sub-vector~$\bs{\theta}^\mu = [\theta_0^\mu,\,\theta_1^\mu]^\top$
devoted to the GEV location.  A similar formula could be used to link
the scale~$\sigma$ (or its log) or the shape~$\xi$ to the
covariates. Note that some model parameters could be shared across the
three GEV parameters. Anyway, it may be of importance to infer on the
trend coefficient~$\theta_1^\mu$.

Using the R package \pkg{NSGEV} \cite{pack_NSGEV} a GEV
model~(\ref{eq:LinEV}) was fitted for the Venice example data with $x$
being the time, along with a constant scale $\sigma$ and a constant
scale $\xi$. The estimate of the trend coefficient $\theta_1^{\mu}$ in
metres per 100-year was $\widehat{\theta}_1^{\mu} = 0.565$ and the
$95\%$ profile likelihood confidence interval on $\theta_1^{\mu}$ was
found to be $[0.284, \, 0.846]$ by using the constrained optimisation
method. However the scaling of $x$ mattered: using the year time unit
instead of the 100-year unit led to incorrect confidence limits. These
results were checked by using the R package \pkg{extRemes}
\cite{pack_extRemes} which also required some manual tuning in order
to get the profile likelihood limits.

In relation with an EV regression one may want to infer on the
$T$-year return period $\eta(T,\, \m{x})$ which now depends on~$\m{x}$
considered as fixed. In the simple case~(\ref{eq:LinEV}) the parameter
$\theta_0^\mu$ could be replaced by $\eta(T,\, x)$ in a
re-parameterisation, provided that $\theta_0^\mu$ is not involved in
the formula for the GEV scale~$\sigma$ nor in that for the
shape~$\xi$. However in a general framework, a re-parameterisation is
awkward since the same parameter can be involved in several GEV
formulas, possibly in a non-linear fashion. By contrast the
optimisation can be implemented for a fairly general framework.

\subsection{Practical considerations on computing}

\subsubsection*{Optimisation}
For a straightforward application of the optimisation formulation, we can use a
routine for constrained optimisation with a general (smooth) constraint.
For instance, the \pkg{NLopt} library \citep{pack_NLopt} offers a number of
algorithms devoted to such optimisation problems, some of them using the
derivatives of the objective and the constraint. This library can be
used from several programming languages including \Prolang{R}, \Prolang{Python}
and \Prolang{Julia}.  The availability of both these exact derivatives in closed
form turns out to be a desirable thing for the sake of computing time. In the
general case where covariates are used, chain rule must be used to compute the
gradient of the log-likelihood.  For instance in the EV regression framework
above, given $\m{x}_i$ and $y_i$, the vector $\bs{\mu}$ of GEV locations
$\mu_i= \mu(\m{x}_i)$ is related to the parameter by a linear relation
$\bs{\mu} = \m{X}^\mu\bs{\theta}$, so the differentiation is
straightforward. The linearity is not essential here: a non-linear but
differentiable link would be used in the same way.

Although most of the available constrained optimisation routines are
\textit{local}, we should ideally use here a \textit{global} constrained
optimisation. The related global algorithms use a local algorithm, hence require
some tuning at both the global and the local levels~\citep{pack_NLopt}.

In order to find suitable initial values, simple algorithms can be designed on
the basis of the geometrical interpretation. We can find a vector $\bs{\theta}$
satisfying the equality $\ell(\bs{\theta}) = \ell_{\max} - \delta$ by using some
kind of dichotomy using both points inside and outside of the likelihood
contour~$\partial \mathcal{R}$. The ML estimate $\hat{\bs{\theta}}$ can be
considered as known, and we can easily find a point
outside~$\partial \mathcal{R}$ with a line search. Provided that such an
algorithm is available in a reliable implementation we can use an equality
constraint rather than an inequality constraint for the optimisation.

Our experience is that even apparently simple situations may be
surprisingly difficult to cope with. Convergence may be difficult to
reach with~$p$ as small as $3$ and with a linear function such as
$\eta(\bs{\theta})= \theta_1$. Another practical difficulty is to make
sure that the convergence was really achieved, which seems more
difficult for a constrained optimisation than for an unconstrained
one. In the second case, the stopping rule will involve both the
change in the objective and that in the constraint, either in absolute
or relative variation. Tuning these parameters can be tedious. In the
classical zero-finding approach for profile likelihood, a graphical
control is most often required since the optimisation used to evaluate
the profile log-likelihood may have failed and/or the zero-finding may
also have failed. Graphical diagnostics such as a plot of the profile
likelihood against the parameter of interest are widely used. These
diagnostics also help for the constrained optimisation method, yet if
the profile likelihood function is to be repeatedly evaluated, the the
constrained optimisation has little benefit.

\subsubsection*{Using MCMC iterates}
\label{SecMCMC}
Programs devoted to MCMC sampling, a.k.a. \textit{MCMC samplers} such as
\pkg{JAGS}~\citep{JAGS} and \pkg{Stan}~\citep{pack_Stan} are widely available
and are fast enough for most purposes.  Over the last decade, considerable
improvements in MCMC sampling have been allowed by the use of Hamiltonian
Monte-Carlo and of automatic differentiation as both implemented in
\pkg{Stan}. Also, some software dedicated to a specific form of models embed
specialised MCMC samplers which can be very fast and over-perform the generalist
samplers.

Although this may not be the simplest and fastest way, profile likelihood
inference can readily and reliably be obtained by using MCMC iterates. Suppose
indeed that on the basis of a prior for $\bs{\theta}$, a large number $K$ of
MCMC iterates~$\bs{\theta}^{[k]}$ have been computed, 
along with the (unnormalised) log-posterior density
$\log p(\bs{\theta}^{[k)}\vert\, \m{y})$ and the log-likelihood
$\ell(\bs{\theta}^{[k]})$. In practice, several thousands of iterates are
typically made available.  These iterates should provide a good coverage of the
region with high posterior probability in the parameter space, and it is a
common practice to consider the Maximum A posteriori Probability (MAP) estimate
as simply being the MCMC iterate for which
$\log p(\bs{\theta}^{[k]}\vert\, \m{y})$ is maximal. Inasmuch a reasonably
non-informative prior for $\bs{\theta}$ is used, the iterates should cover as
well the region with high likelihood and the ML estimate $\hat{\bs{\theta}}$ can
similarly be obtained as the iterate $\bs{\theta}^{[k]}$ for
which~$\ell(\bs{\theta}^{[k]})$ is maximal.

\begin{figure}
  \centering
  \includegraphics[width=0.8\textwidth]{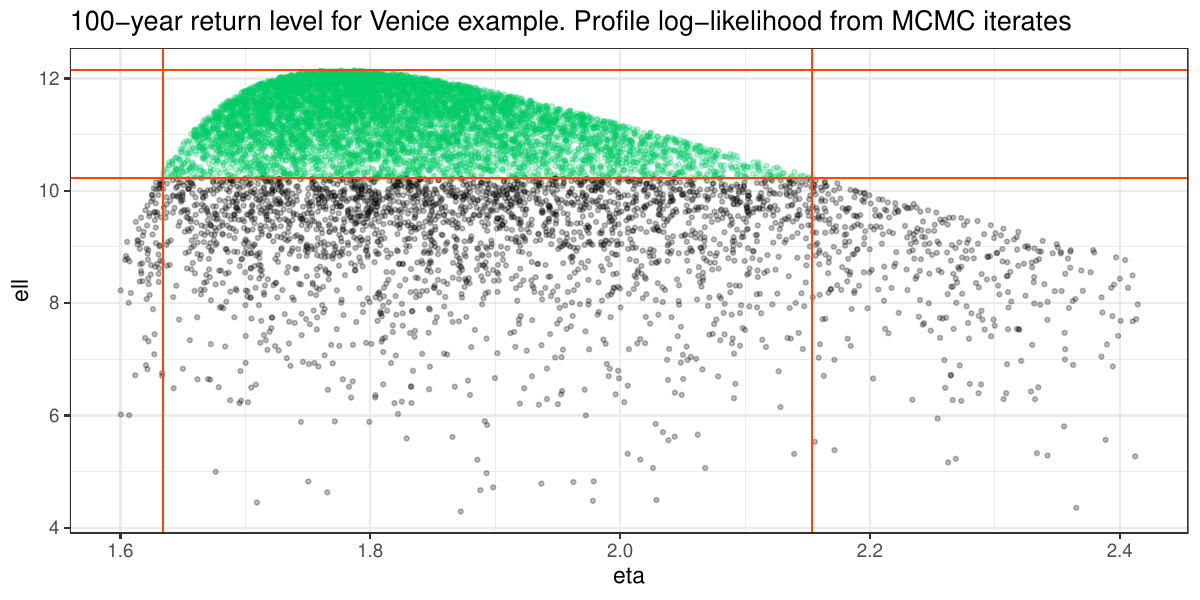}
  \caption{\label{FigMCMC} \sf \footnotesize Profile likelihood interval from
    $K= 10^4$ MCMC iterates. The $95\%$ limits are for the $100$-year return
    level $\eta(T)$ for the GEV example.  The confidence limits (vertical lines)
    are the smallest and the largest $\eta(T)$ corresponding to iterates
    $\bs{\theta}^{[k]}$ such that
    $\ell(\bs{\theta}^{[k]}) \geqslant \ell_{\max} - \delta$. The corresponding
    iterates are shown as green points $[\eta, \, \ell]$. The confidence limits
    are identical to those obtained by constrained optimisation up to two
    decimal places. }
\end{figure}

As recalled before, the Bayesian inference on a function~$\eta(\bs{\theta})$ of
the parameter, be it scalar-valued or vector-valued, is straightforward. If
$\eta(\bs{\theta})$ is scalar-valued, we get an approximation of the profile
likelihood bound~$\Ub{\eta}$ as the maximal value $\eta(\bs{\theta}^{[k]})$
among the iterates $\bs{\theta}^{[k]}$ such that the log-likelihood inequality
$\ell(\bs{\theta}^{[k]}) \geqslant \ell_{\max} - \delta$ holds.
Although not required, the profile log-likelihood function
$\Prof{\ell}(\eta)$ can also in this framework be estimated as the
maximum of the log-likelihoods $\ell(\bs{\theta}^{[k]})$ over the
iterates such that $\eta(\bs{\theta}^{[k]})$ is close to $\eta$. This
is illustrated with the Venice example on Figure~\ref{FigMCMC} for the
$100$-year return level $\eta(100)$. The MCMC iterates were obtained
by using the very efficient \verb@rpost_rcpp@ function of the
\pkg{revdbayes} R package~\citep{pack_revdbayes}; The
\code{"flatflat"} GEV prior was used. For EV regression models with
covariates the \pkg{Stan} sampler can be used.

\section{Function with an extra variable and ODE}

\label{SecExtraODE}

\subsection{Problem}
In a number of situations we want to infer on a smooth function
$\eta(\bs{\theta},\,t)$ where $t$ is an extra continuous variable assumed to be
scalar. An iconic example is provided by the simple linear regression
$y = \theta_0 + \theta_1 t + \varepsilon$ where $t$ is a single covariate
and $\varepsilon$ is a noise term. This example extends to non-linear regression
models with a single covariate and Gaussian noise~\citep{BatesWatts}. Another
common example is when $\eta(\bs{\theta},\,t)$ is a probability function such as
the cumulative distribution or the quantile function, $t$ being then a quantile
or a probability. For the GEV return level problem above, we can take $t$ to be
the return period~$T$, or some strictly monotonous function of~$T$. In such a
framework, the confidence interval for $\eta$ will typically be required for
\textit{several} values of~$t$ in order to form a confidence band as illustrated
with the return level plot above on Figure~\ref{FigVeniceRLPlot}.  For consistency with the examples, we
will consider the notations $\eta(\bs{\theta},\,t)$ and $\eta(t;\,\bs{\theta})$
as equivalent. It will also be convenient to refer to~$t$ as the \textit{time}
variable.

\subsection{From optimisation to ODE}
\label{Optim2ODE}
Going back to the optimisation problem discussed above, the objective now
depends on~$t$, while the constraint does not. The parameter value
$\Ub{\bs{\theta}}(t)$ corresponding to the upper confidence bound --\,now
depending on~$t$\,-- remains on the same log-likelihood
contour~$\partial \mathcal{R}$ as before. Provided that the
function~$\eta(\bs{\theta},\,t)$ is smooth, the parameter $\Ub{\bs{\theta}}(t)$
moves along a smooth path on $\partial \mathcal{R}$.  In the general case where
an explicit solution of the optimisation problem does not exist, we can solve as
many optimisation problems as there are wanted values~$t_i$ for~$t$.  Since
$\Ub{\bs{\theta}}(t)$ smoothly depends on~$t$, the solution for a time $t = t_i$
can be used as initial value for the next time~$t = t_{i+1}$ or $t = t_{i-1}$,
provided that the sequence $t_i$ is monotonous. This leads to some savings in
the number of iterations, especially if the $t_i$ are close to each other.  A
natural idea to avoid solving many optimisation problems is to find an ODE
describing the evolution of $\Ub{\bs{\theta}}(t)$. Roughly speaking, the ODE
comes by differentiating the KKT conditions with respect to the time~$t$ as we
now explain.

We will use the dot notation for the derivative w.r.t. the time~$t$, as in
$\Dot{\Ub{\bs{\theta}}}(t)$ for the derivative of $\Ub{\bs{\theta}}(t)$ and also
use $\Ub{\eta}(t) := \eta\{\Ub{\bs{\theta}}(t), \,t\}$. Similarly,
$\Ub{\nu}(t)$ will be the Lagrange multiplier used to find~$\Ub{\bs{\theta}}(t)$
from the KKT conditions~(\ref{eq:Lagrangea}) and~(\ref{eq:Lagrangeb}) with
$\eta(\bs{\theta})$ replaced by $\eta(\bs{\theta}, \,t)$
in~(\ref{eq:Lagrangea}). Since $\Ub{\bs{\theta}}(t)$ moves on the boundary
$\partial \mathcal{R}$, the derivative vector $\Dot{\Ub{\bs{\theta}}}(t)$
remains orthogonal to the hyperplane tangent to $\mathcal{R}$ at
$\Ub{\bs{\theta}}(t)$. This can be stated by differentiating the
equation~(\ref{eq:Lagrangeb}) w.r.t.~$t$, which gives
\begin{equation}
  \label{eq:tangent}
  \frac{\partial \ell}{\partial \bs{\theta}^\top} \, \Dot{\Ub{\bs{\theta}}}  = 0,
\end{equation}
where the partial derivative is evaluated at
$\Ub{\bs{\theta}}(t)$. We can similarly differentiate w.r.t.~$t$ the
equation~(\ref{eq:Lagrangea}) where $\bs{\theta}$ and $\nu$ are
replaced by the solutions $\Ub{\bs{\theta}}(t)$ and
$\Ub{\nu}(t)$, leading to
\begin{equation}
  \label{eq:ODE0}
   \frac{\partial^2 \eta}{\partial t \partial \bs{\theta}} + 
   \frac{\partial^2 \eta}{\partial \bs{\theta} \partial \bs{\theta}^\top} \,
   \Dot{\Ub{\bs{\theta}}} = 
   \Dot{\Ub{\nu}}
   \, \frac{\partial \ell}{\partial \bs{\theta}}
   +  \Ub{\nu}
   \frac{\partial^2 \ell}{\partial \bs{\theta} \partial \bs{\theta}^\top}\,
   \Dot{\Ub{\bs{\theta}}},
 \end{equation}
 and by rearranging, (\ref{eq:tangent}) and (\ref{eq:ODE0}) can be written as
 \begin{equation}
   \label{eq:ODEp1}
   \begin{bmatrix}
     - \partial^2_{\bs{\theta},\bs{\theta}^\top} \eta
     +  \Ub{\nu} \, \{\partial^2_{\bs{\theta},\bs{\theta}^\top} \ell \}
     &
     \partial_{\bs{\theta}} \ell \\
     \partial_{\bs{\theta}^\top} \ell & 0 \rule{0pt}{1.2em}
   \end{bmatrix}
   \begin{bmatrix}
     \Dot{\Ub{\bs{\theta}}} \\ \Dot{\Ub{\nu}} \rule{0pt}{1.2em}
   \end{bmatrix}
   =
   \begin{bmatrix}
     \partial^2_{t, \bs{\theta}} \eta  \\
     \rule{0pt}{1.2em} 0
  \end{bmatrix}.
\end{equation}
Provided that the $(p + 1) \times (p +1)$ matrix of the left-hand side is
invertible, we get an ODE for the vector
$\Ub{\bs{\theta}}^\dag := [\Ub{\bs{\theta}}^\top,\, \Ub{\nu}]^\top$.  As will be
illustrated later, in some cases $\Ub{\nu}$ can be expressed as a function of
$\Ub{\bs{\theta}}$ and we then get an ODE
\begin{equation}
  \label{eq:ODE}
  \Dot{\bs{\theta}}  = \Ub{\m{b}}(\bs{\theta},\,t) 
\end{equation}
where $\Ub{\m{b}}$ stands for a vector-valued function with arguments
$\bs{\theta}$ and $t$. Anyway, using the ODE requires the availability of the
two Hessians for the objective $\eta$ and for the log-likelihood, and also the
availability of the~$p$ cross derivatives $\partial^2_{t,\theta_k} \eta$.

\begin{rk}
  When $\eta(\bs{\theta},\,t)$ is the value of a cumulative distribution
  function $F(t;\,\bs{\theta})$, the cross derivatives
  $\partial^2_{t,\theta_k} \eta$ are simply the first-order derivatives
  $\partial_{\theta_k} f$ of the density function~$f$, which can be computed
  from the gradient of the log-likelihood.
\end{rk}

\begin{rk}
  \label{IncreasingBounds}
  Note that with partial derivatives evaluated at $\Ub{\bs{\theta}}(t)$ and $t$,
  we have
  $$
  \Dot{\Ub{\eta}} = \frac{\partial \eta}{\partial t} + 
  \frac{\partial \eta}{\partial \bs{\theta}^\top} \Dot{\Ub{\bs{\theta}}}
  = \frac{\partial \eta}{\partial t} + \Ub{\nu}^{-1}
  \frac{\partial \ell}{\partial \bs{\theta}^\top} \Dot{\Ub{\bs{\theta}}} 
  = \frac{\partial \eta}{\partial t},
  $$
  because $\Dot{\Ub{\bs{\theta}}}$ is in the hyperplane tangent to
  $\partial \mathcal{R}$ hence~(\ref{eq:tangent} holds. A similar
  relation holds for the lower bound~$\Lb{\eta}(t)$. This proves that
  if $\eta(\bs{\theta},\,t)$ is increasing w.r.t. $t$ for any
  admissible~$\bs{\theta}$, then both bounds $\Lb{\eta}$ and
  $\Ub{\eta}$ will be increasing w.r.t.~$t$, which seems a natural and
  appealing feature. By contrast, the lower bound given by the delta
  method may fail to be increasing.
\end{rk}

\subsection{GEV return levels}
\label{GEVt}
We now give some details on the ODE derivation for the inference on the return
level $\eta(\bs{\theta},\,t)$ corresponding to the return period~$t$ and related
to the GEV distribution with parameter
$\bs{\theta} = [\mu,\,\sigma,\,\xi]^\top$. The return level was defined
by~(\ref{eq:RL}) and the problem was discussed in
section~\ref{SecGEVRL}. Because $\partial_\mu \eta = 1$, we can use the first
coordinate of the vector-valued
equation~$\partial_{\bs{\theta}} \eta = \nu \, \partial_{\bs{\theta}} \ell$ to
eliminate the Lagrange multiplier $\nu$ from the equations for the two remaining
coordinates, getting
\begin{equation}
  \label{eq:SigmaXi}
  \left\{
    \begin{array}{r c l}
      \partial_\sigma \eta &=
      & \partial_\sigma \ell \, \left\{\partial_\mu  \ell\right\}^{-1}\\
      \partial_\xi \eta &=
      & \partial_\xi \ell \, \left\{\partial_\mu  \ell\right\}^{-1}. 
        \rule{0pt}{1.2em}
    \end{array}
\right.
\end{equation}
Let $r_\sigma$ and $r_\xi$ be the two functions of $\bs{\theta}$ defined by the
right-hand sides of the two last equations.

Let~$\bs{\theta}(t)$ be such that the KKT conditions (\ref{eq:Lagrangea})
and (\ref{eq:Lagrangeb}) hold for any~$t$. Then~(\ref{eq:SigmaXi}) also holds
for any~$t$. Note that $r_{\sigma}\{\bs{\theta}(t)\}$ and
$r_{\xi}\{\bs{\theta}(t)\}$ only depend on~$t$ through $\bs{\theta}(t)$. By
differentiating w.r.t.~$t$ we get the following next two equations
\begin{equation}
  \label{eq:ODEGEV}
  \arraycolsep=1.4pt\def\arraystretch{1.2}
  \left\{
    \begin{array}{r c r c r c r}
      \left[\partial^2_{\mu,\sigma} \eta 
        - \partial_\mu  r_\sigma \right]  \Dot{\mu} &+&
      \left[\partial^2_{\sigma,\sigma}\eta  
        - \partial_\sigma r_\sigma \right]  \Dot{\sigma} &+&
      \left[\partial^2_{\xi,\sigma} \eta 
        -  \partial_\xi  r_\sigma \right]  \Dot{\xi} &=&
      -\partial^2_{t,\sigma} \eta,  \\
      \left[\partial^2_{\mu,\xi} \eta 
        - \partial_\mu  r_\xi \right]  \Dot{\mu} &+&
      \left[\partial^2_{\sigma,\xi}\eta  
        - \partial_\sigma r_\xi \right]  \Dot{\sigma} &+&
      \left[\partial^2_{\xi,\xi} \eta
        -  \partial_\xi  r_\xi \right]  \Dot{\xi} &=&
      -\partial^2_{t,\xi} \eta, \\
      \partial_\mu \ell  \: \Dot{\mu} &+& 
      \partial_\sigma \ell \: \Dot{\sigma} &+& 
      \partial_\xi \ell  \: \Dot{\xi} &=& 0. 
    \end{array}\right.
\end{equation}
The third equation is a simple translation of the vector
form~(\ref{eq:tangent}). By solving the linear system~(\ref{eq:ODEGEV}) with
unknown $\Dot{\bs{\theta}}$, we get the form~(\ref{eq:ODE}) above.


The ODE formulation provides nice hints about the confidence bounds.
\begin{itemize}
\item When the return period $t$ tends to $\infty$, we may conjecture that the
  value of $\Ub{\bs{\theta}}(t)$ tends to the GEV parameter~$\bs{\theta}$ that
  maximises the shape parameter $\xi$ for~$\bs{\theta} \in \mathcal{R}$, or
  equivalently for $\bs{\theta} \in \partial \mathcal{R}$. This is due to the
  fact that for large~$t$ the return level depends on $\xi$ more heavily than on
  $\mu$ and $\sigma$.
  
\item It is easily checked that $\partial_t \eta(\bs{\theta}, \,t) \geqslant 0$
  for any admissible GEV parameter $\bs{\theta}$. So both the upper and the
  lower bounds of the confidence interval are increasing with~$t$, see
  Remark~\ref{IncreasingBounds} above.
\end{itemize}

\begin{figure}
   \centering
   \includegraphics[width=0.55\textwidth]{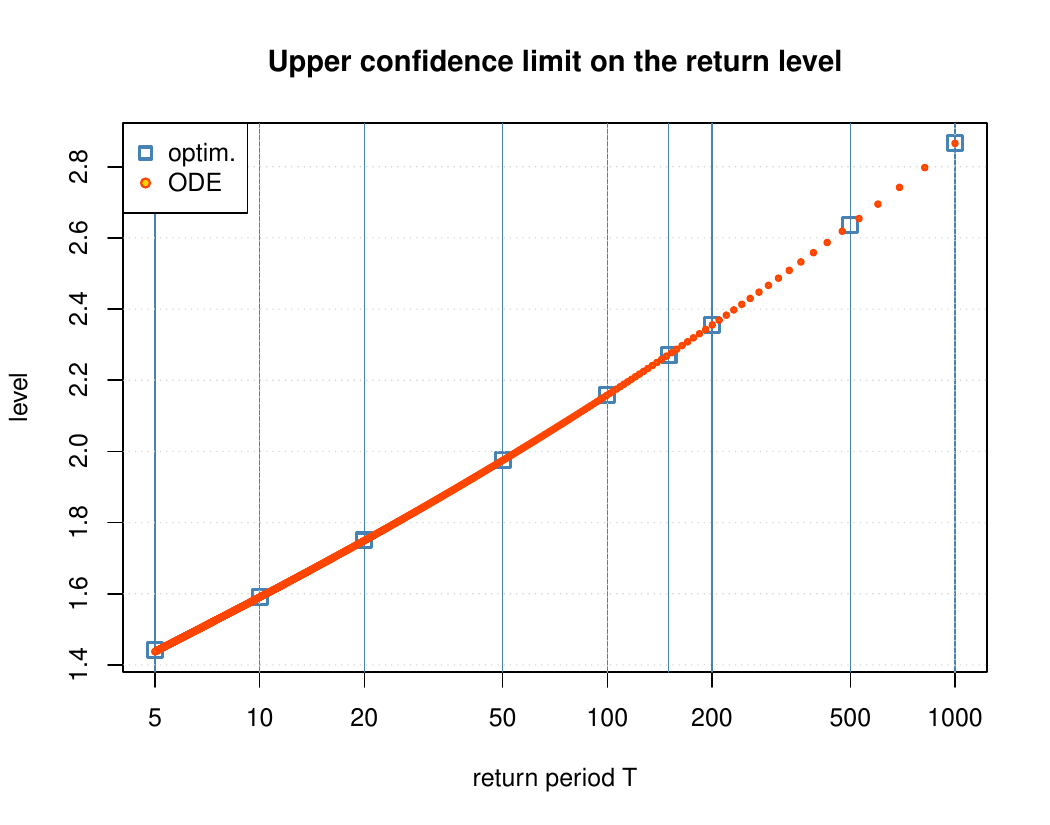}
   \includegraphics[width=0.4\textwidth]{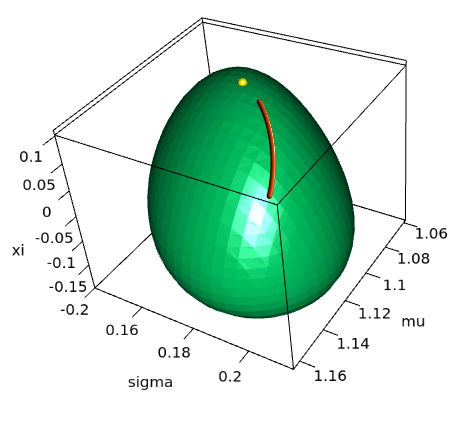}
   \caption{\label{FigODE} \footnotesize Confidence interval for the return
     levels $\eta(t)$ of the Venice example via ODE. Left: The upper bounds
     $\Ub{\eta}(t)$ computed by solving the ODE (orange bullets) and those
     obtained by constrained optimisation (blue squares). Right: The parameters
     $\Ub{\bs{\theta}}(t)$ shown as orange spheres on the
     surface~$\partial \mathcal{R}$ in the parameter $\mu\sigma\xi$-space. The
     yellow sphere on the very top of the surface shows the value $\bs{\theta}$
     that maximises~$\xi$, the corresponding coordinate~$\xi$ being the upper
     bound $\Ub{\xi}$ of the confidence interval on $\xi$.}
 \end{figure}

The ODE method is illustrated for the Venice example on Figure~\ref{FigODE}.
The left part shows the upper confidence limit $\Ub{\eta}(T)$ for several return
periods~$T$. The right part shows the corresponding path $\Ub{\bs{\theta}}(T)$
on the likelihood contour~$\partial \mathcal{R}$. The function \code{lsoda} from
the R package \pkg{deSolve} \citep{pack_deSolve} was used to compute the
numerical solution of the ODE using $1000$ discrete times corresponding to
unevenly spaced return periods. The initial value $\Ub{\bs{\theta}}(0)$ for
$t = 0$ was obtained by constrained optimisation.
 
 \section{ODEs and contours}
\label{SecODEContours}

\subsection{Profile contours}
\label{SecProfileContour}
We now consider the case where the parameter of interest is no longer scalar but
is a vector $\bs{\psi}$ with length~$d$, and as before
$\bs{\theta} = [\bs{\psi}^\top, \, \bs{\lambda}^\top]^\top$.  A $d$-dimensional
approximate confidence region with level $1-\alpha$ for $\bs{\psi}$ is defined
by
\begin{equation}
  \label{eq:Profd}
  \Prof{\ell}(\bs{\psi}) \leqslant \ell_{\max}- \delta, \qquad
  \delta = q_{\chi^2(d)}(1 - \alpha) /2.
\end{equation}
Corresponding to the case where the inequality is an equality, the boundary of
the region in the $\bs{\psi}$-space defined by~(\ref{eq:Profd}) may be called a
\textit{profile likelihood contour}, or simply: a \textit{profile contour}. Of
special interest is the case $d=2$: the profile contour will generally be a
smooth Jordan curve, the interior being the confidence region.  We will not try
to give conditions ensuring that this kind of favourable situation holds.

The profile contour can be explored by using constrained optimisations. In order
to get points $\bs{\psi}$ on the contour we can indeed optimise some well-chosen
scalar functions of the parameter $\bs{\theta}$ constrained to lie in the
$p$-dimensional region $\mathcal{R}(\delta)$ of~(\ref{eq:Rdelta}), with the
value of~$\delta$ related to the length~$d$ of the parameter of interest
according to~(\ref{eq:Profd}). For a given vector $\m{a}$ with length $d$, it is
intuitively clear that by maximising or minimising
$\eta(\bs{\theta}) := \m{a}^\top \bs{\psi}$ for $\bs{\theta}$ in
$\mathcal{R}(\delta)$, we get a point $\bs{\theta}^\star$ such that
$\bs{\psi}^\star:= \bs{\psi}(\bs{\theta}^\star)$ lies on the contour. Moreover,
the hyperplane tangent to the contour at $\bs{\psi}^\star$ is orthogonal
to~$\m{a}$.

As a further step, we can consider a fixed family of vectors $\m{a}(t)$ smoothly
depending on a real quantity~$t$, say $0 \leqslant t \leqslant t^\star$, in
order to get a continuum of points $\bs{\psi}^\star(t)$ on the contour. We will
further assume that $\m{a}(t)$ has an unit Euclidean norm $\| \m{a}(t)\| = 1 $
for all~$t$. The first-order condition is
\begin{equation}
  \label{eq:ContourOrd1}
  \begin{bmatrix}
    \m{a}(t) \\
    \m{0}
  \end{bmatrix}
  = \nu \,\frac{\partial \ell}{\partial \bs{\theta}},
\end{equation}
where $\nu$ is a Lagrange multiplier depending on~$t$ and the vector of
zeros~$\m{0}$ has length $p - d$. Since $\m{a}(t)$ has unit norm, we find
$\nu = \pm u(t)^{-1/2}$, where $u(t)$ is defined as the squared norm of the
score vector $\partial_{\bs{\theta}} \ell$ evaluated at $\bs{\theta}(t)$. We
will now on consider only the ``plus sign'' case $\nu = u(t)^{-1/2}$. It will
help to use notations for the score vector and the negative Hessian
\begin{equation}
  \label{eq:ODEContour2}
  \m{z} := \frac{\partial \ell}{\partial \bs{\theta}}, \quad
  \m{H} :=  - \frac{\partial^2 \ell}{\partial \bs{\theta} \partial \bs{\theta}^\top},
\end{equation}
reminding that these quantities depend on $t$ through $\bs{\theta}(t)$. We have 
\begin{equation}
  \label{eq:uuDot}
  u(t) = \m{z}^\top\m{z} \quad \text{and} \quad
  \Dot{u}(t) = - \Dot{\bs{\theta}}^\top \m{H} \,\m{z} -
  \m{z}^\top \, \m{H} \, \Dot{\bs{\theta}}.
\end{equation}
By differentiating the first-order condition~(\ref{eq:ContourOrd1}) w.r.t.~$t$
we get
\begin{equation*}
  \begin{bmatrix}
    \Dot{\m{a}}(t) \\
    \m{0}
  \end{bmatrix} = 
  \Dot{\nu}
  \, \frac{\partial \ell}{\partial \bs{\theta}}
  + \nu
  \frac{\partial^2 \ell}{\partial \bs{\theta} \partial \bs{\theta}^\top}\,
  \Dot{\bs{\theta}},
\end{equation*}
where the partial derivatives are in both cases evaluated
at~$\bs{\theta}(t)$. Using the fact that $\Dot{\nu} = -2^{-1}\Dot{u}\,u^{-3/2}$
where $u$ and $\Dot{u}$ are given by~(\ref{eq:uuDot}), we get after rearranging
\begin{equation}
  \label{eq:ODEContour}
  \begin{bmatrix}
    \Dot{\m{a}} \\
    \m{0}
  \end{bmatrix} = 
   - u^{-1/2} \,
  \left\{\m{I}_p - \frac{\m{z}\m{z}^\top}{\m{z}^\top\kern-0.2em\m{z}} \right\}
  \m{H} \, \Dot{\bs{\theta}} = \m{B} \,\dot{\bs{\theta}}.
\end{equation}
In this expression, the matrix between the curly brackets is clearly that of
the projection on the orthogonal supplement $\m{z}^\perp$ of the linear subspace
spanned by $\m{z}$ in $\mathbb{R}^p$. So $\m{B}$ has rank~$p-1$ and therefore we
can not find $\Dot{\bs{\theta}}$ from the sole
equation~(\ref{eq:ODEContour}). However we have one more equation for
$\Dot{\bs{\theta}}$, namely $\m{z}^\top \Dot{\bs{\theta}} = 0$ and we can get
$\Dot{\bs{\theta}}$ as a least squares solution e.g., by using a Housholder QR
decomposition of the matrix obtained by adding the new row $\m{z}^\top$ to
$\m{B}$. Note that the left-hand side of~(\ref{eq:ODEContour}) must be
orthogonal to~$\m{z}$ and the solution of the linear system of $p+1$ equations
is then \textit{exact}, so the equation defines an ODE for $\bs{\theta}(t)$.


By numerically solving the ODE~(\ref{eq:ODEContour}) above, we will find a
continuum of points on the contour.  As before, we need an initial value
$\bs{\theta}(0)$ which can be obtained by optimisation or by some alternative
method as the one described in the next section. In the first case, to account
for a possible failure of the constrained optimisation, a few optimisations may
be needed to get the initial value. Solving then the ODE is then very fast.

\begin{figure}
  \centering
  \begin{tabular}{c c}
    \includegraphics[width=0.6\textwidth]{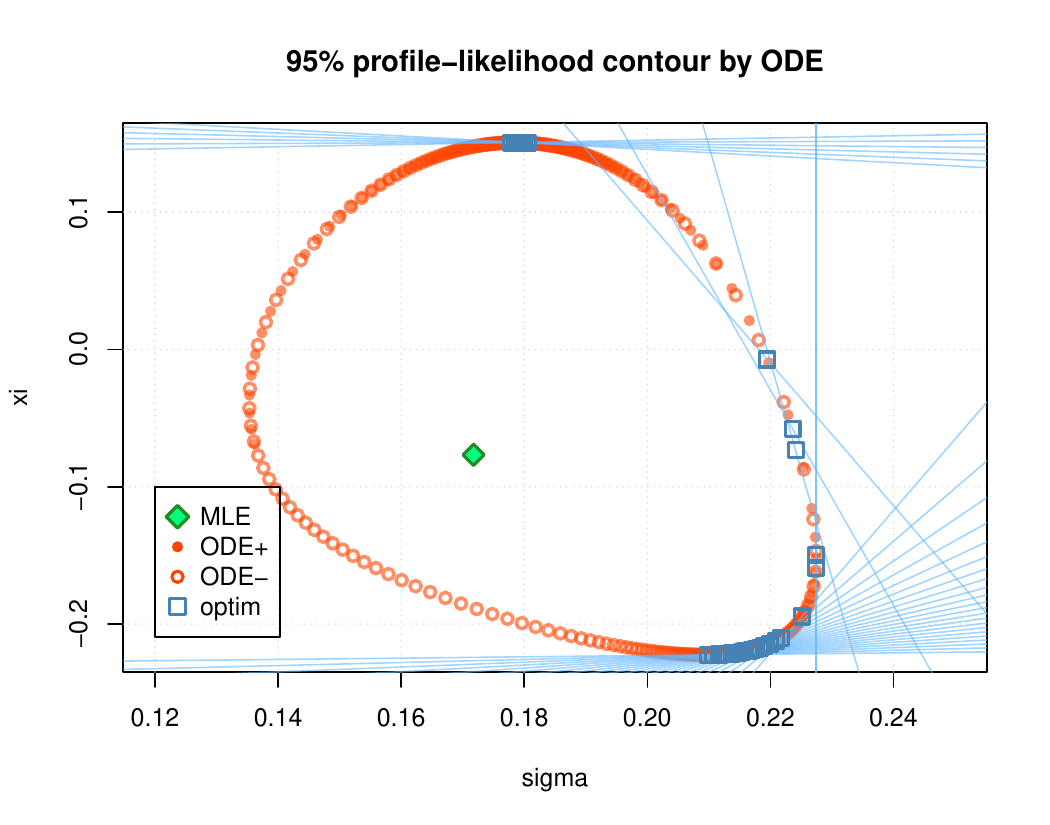}
    & \raisebox{2em}{\includegraphics[width=0.35\textwidth]{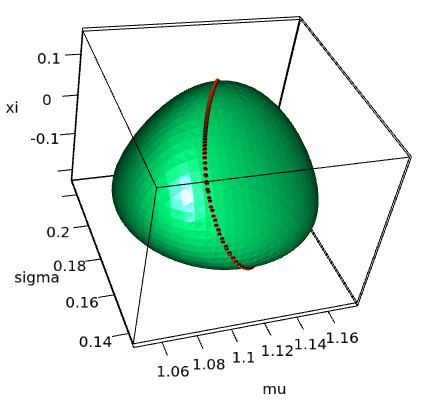}}
  \end{tabular}
  \caption{\label{VeniceContour} \footnotesize \sf Profile likelihood contour
    for the GEV parameters $\sigma$ and $\xi$ for the Venice example.  Left:
    Profile contour. Right: the path $\bs{\theta}(t)$ corresponding to the
    solution of the ODE. The vector $\bs{\theta}(t)$ moves on the boundary of
    the high-likelihood region $\mathcal{R}(\delta)$. Remind that the value of
    $\delta$ used here corresponds to $d=2$ d.f. hence differs from the one used
    for the confidence intervals in Figure~\ref{FigGEV}, corresponding to one
    d.f..}
\end{figure}

The use of~(\ref{eq:ODEContour}) is illustrated on Figure~\ref{VeniceContour}
for the Venice example and for the couple of
interest~$\bs{\psi} := [\sigma, \, \xi]^\top$.  The left panel shows points on
the contour obtained by constrained optimisation and by ODE. For the
optimisation, a large number of directions $\m{a}$ were tested; When the
optimisation succeeded the corresponding point is shown as a square, along with
the tangent which by construction must have the prescribed direction. Note that
although each square corresponds to an optimisation reported as successful, some
problems can still be seen on the right part of the contour: one point is
slightly out of the contour and another one does not have the prescribed
tangent. For the ODE, we used again the R package~\pkg{deSolve} and the vectors
$\m{a}(t) := [\cos(t), \,\sin(t)]^\top$ with discrete times evenly spaced
between~$0$ and $2 \pi$. The result of a successful optimisation was used as the
initial value $\bs{\theta}(0)$. There are actually two ODE solutions,
corresponding to the choice of sign in $\nu = \pm u^{-1/2}$. Both ODE solutions
are required to cover the whole contour and the solution paths partially
overlap, yet very consistently.  The two paths $\bs{\theta}(t)$ on
$\partial \mathcal{R}(\delta)$ are shown in the right panel of
Figure~\ref{VeniceContour}. An observer located in the parameter space far away
in the direction of the $\mu$-axis would see the paths as shown in left panel,
the $\sigma$ and the $\xi$ axes then becoming the $x$ and $y$ axes in the usual
meaning.

\subsection{Log-likelihood contours and confidence level}
Turning back to the inference on a smooth scalar function $\eta(\bs{\theta})$ of
the parameter, the ML estimate~$\hat{\bs{\theta}}$ can be used to derive an
initial value for a constrained optimisation algorithm. Yet a possibly better
use of~$\hat{\bs{\theta}}$ can be made.  In the likelihood contour
$\partial \mathcal{R}(\delta)$, the quantity $\delta \geqslant 0$ giving the
(approximate) confidence level can be regarded as a continuous variable for the
optimisation constraint. The variation of $\delta$ should induce a smooth
deformation of the likelihood contour $\partial \mathcal{R}(\delta)$. For
$\delta \approx 0$ the likelihood region is a small ellipsoid
with centre~$\hat{\bs{\theta}}$, and by increasing $\delta$ we get
some kind of ``inflating bubble''. If we consider both the upper confidence
limit $\Ub{\eta}$ and the related parameter vector $\Ub{\bs{\theta}}$ as
depending on $\delta$, we can derive an ODE for
$\Ub{\bs{\theta}}^\dag := [\Ub{\bs{\theta}}^\top, \, \Ub{\nu}]^\top$ where
$\Ub{\nu}$ is the Lagrange multiplier. Unlike to what occurred before, the vector
$\Ub{\bs{\theta}}(\delta)$ no longer moves on a given likelihood contour, but
instead remains on the surface of the bubble as it goes inflating. The derivation
hence slightly differs from what was described above.

\begin{rk}
  This method is similar to that described
  by~\cite{StaporEtAl_OptimizationProfile}. However the authors use the
  log-likelihood as objective and (in our notations) the function
  $\eta(\bs{\theta})$ to define an equality constraint as
  in~(\ref{eq:OptimConstrPsi}) for the special case $\eta(\bs{\theta}) =\psi$.
\end{rk}

To stick to the notations used in the previous sections, we will use the
symbol~$t$ in place of $\delta$ in the rest of this section. We can derive an
ODE for $\Ub{\bs{\theta}}^\dag(t)$ by following nearly the same steps as those
described in section~\ref{Optim2ODE}. As a major difference, in the constrained
optimisation the objective $\eta$ no longer depends on~$t$, but the constraint
now does.  By differentiating~(\ref{eq:Lagrangeb}) w.r.t. $t$ we now get
$$
\frac{\partial \ell}{\partial \bs{\theta}^\top} \, \Ub{\Dot{\bs{\theta}}} = - 1,
$$
and the equivalent of (\ref{eq:ODEp1}) for the new framework is now
\begin{equation}
  \label{eq:ODEp1Bubble}
  \begin{bmatrix}
    - \partial^2_{\bs{\theta},\bs{\theta}^\top} \eta
    +  \Ub{\nu} \, \{\partial^2_{\bs{\theta},\bs{\theta}^\top} \ell \}
    &
    \partial_{\bs{\theta}} \ell \\
    \partial_{\bs{\theta}^\top} \ell & 0 \rule{0pt}{1.2em}
  \end{bmatrix}
  \begin{bmatrix}
    \Dot{\Ub{\bs{\theta}}} \\ \Dot{\Ub{\nu}} \rule{0pt}{1.2em}
  \end{bmatrix}
  =
   \begin{bmatrix}
     \m{0}  \\
     \rule{0pt}{1.2em} -1
  \end{bmatrix}. 
\end{equation}
As for the initial condition, we could think of taking
$\Ub{\bs{\theta}}(0) := \hat{\bs{\theta}}$. Yet the initial Lagrange multiplier
$\Ub{\nu}(0)$ is now undetermined because the gradient of the log-likelihood
$\partial_{\bs{\theta}} \ell$ is zero at $\hat{\bs{\theta}}$. However we can
instead start from an approximation of the value
$\Ub{\bs{\theta}}^\dag(\delta_1)$ corresponding to a small time $\delta_1 >
0$. For a small enough value, the contour of the log-likelihood is approximately
an ellipsoid and the function $\eta(\bs{\theta})$ is approximately linear
inside the contour. By reasoning as we did in the linear regression example of
Section~\ref{SecLinReg}, we get the following approximations for $\nu(\delta_1)$
and $\bs{\theta}(\delta_1)$
$$
 \Ub{\widetilde{\nu}}(\delta_1) := \{\m{h}^\top_0 \m{H}_0^{-1}\m{h}_0\}^{1/2} / \sqrt{2 \delta_1},
 \qquad  \Ub{\widetilde{\bs{\theta}}}(\delta_1) := \hat{\bs{\theta}} +
 \Ub{\widetilde{\nu}}^{-1}(\delta_1) \, \m{H}^{-1}_0 \m{h}_0,
$$
where $\m{H}_0$ is the Hessian of the negative log-likelihood and
$\m{h}_0$ is the gradient of $\eta(\bs{\theta})$, both taken for
$\bs{\theta} = \hat{\bs{\theta}}$.  The matrix $\m{H}_0$ should be
positive definite. By changing the sign of the initial Lagrange
multiplier, the same ODE can be used to find $\Lb{\bs{\theta}}(t)$
corresponding to the lower confidence bound.

It may happen that the solution of~(\ref{eq:ODEp1Bubble}) with the given initial
conditions does not exist on the full interval $(0,\, \delta)$ and only exists
on a smaller interval, corresponding to a confidence level higher than the one
wanted. The region $\mathcal{R}(\delta)$ may indeed fail to be connected, and
this can be revealed by the solution of the ODE.

\begin{rk}
  To a certain extend, solving the ODE (\ref{eq:ODEp1Bubble}) is the opposite of
  maximising the log-likelihood: this somewhat ``undoes'' what the maximisation
  did. Considered in reverse time, the ODE admits the equilibrium point
  $\hat{\bs{\theta}}$ which is Lyapunov-stable since the log-likelihood is by
  construction a Lyapunov function.
\end{rk}

\begin{figure}
  \centering
  \begin{tabular}{c c}
    \includegraphics[width=0.5\textwidth]{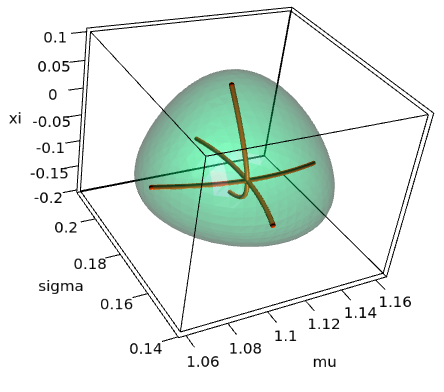}
    & \includegraphics[width=0.5\textwidth]{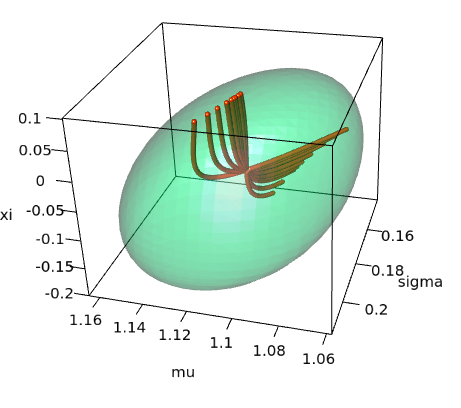}
  \end{tabular}
  \caption{\label{VeniceODEProfile} \footnotesize \sf Likelihood contour and
    paths corresponding to the solution of the ODEs for the Venice
    example. Left: Six paths corresponding to the determination of lower and
    upper confidence bounds for the three GEV parameters. Right: Paths
    corresponding to the lower and upper bounds for nine return periods. In both
    cases each path starts from a point close to the ML estimate
    $\hat{\bs{\theta}}$ and ends at a point on the contour which is shown as the
    semi-transparent green surface.}
\end{figure}

  
The method is illustrated with the Venice example on
Figure~\ref{VeniceODEProfile}. The left part shows the paths used to get a
confidence interval on each of the three GEV parameters $\mu$, $\sigma$ and
$\xi$.  For each parameter~$\theta_k$, two paths are obtained starting from a
point close to $\hat{\bs{\theta}}$ and ending at a point
$\Lb{\bs{\theta}}^{[k]}$ or $\Ub{\bs{\theta}}^{[k]}$ on the contour of the
log-likelihood $\partial \mathcal{R}$ such that the parameter of interest
$\theta_k$ is either minimised or maximised. The right part shows the
determination of the upper bound on the return period $\eta(T)$. Nine return
periods $T_i$ were fixed and for each of them we proceeded as we did for the GEV
parameters. In both cases, the paths corresponding to the minimisation and the
maximisation of a same quantity are parts of a smooth path that starts from
$\Lb{\bs{\theta}}$ and ends at the corresponding $\Ub{\bs{\theta}}$.

Still using the Venice data, a linear time trend was used in an EV
regression as in (\ref{eq:LinEV}). As was the case with the
optimisation method, a suitable scaling was required to get the exact
confidence bounds: the unit for the time covariate $x$ had to be set
to $100$-year. Interestingly, plotting the Lagrange multiplier $\nu$
against the time of the ODE may provide a valuable diagnostic. The
value $\nu(t)$ seems to tend to a limit for large $t$ and when the
parameter is not suitably scaled the limit is reached in a few time
steps.

\begin{rk}
  It is easy to see from (\ref{eq:ODEp1Bubble}) that when $\eta$ is a
  linear function of $\bs{\theta}$ the Lagrange multiplier $\nu(t)$ is
  a monotonous function of the time. The $2$-nd order derivative of
  $\eta$ then cancels. 
\end{rk}

The method described here is both very simple and fairly general. It
can be used to provide initial values for another ODE
computation as introduced in Sections~\ref{GEVt}
and~\ref{SecProfileContour}, a constrained optimisation then no longer
being required. For instance, for the GEV return levels we could find
a path from $\hat{\bs{\theta}}$ to the vector $\Ub{\bs{\theta}}^{[j]}$
corresponding to the maximisation of $\eta(T_j)$ for some return
period~$T_j$, and then solve an ODE with a time related to the return
period, as discussed in Section~\ref{GEVt}. In other words, we could
move from the point $\hat{\bs{\theta}}$ inside the region
$\mathcal{R}(\delta)$ to a point $\Ub{\bs{\theta}}^{[j]}$ on the
boundary and then move on the boundary, rather than repeatedly move
from $\hat{\bs{\theta}}$ to a boundary point as shown on the right
side of Figure~\ref{VeniceODEProfile}. The first option should be
faster, but using both these two fast approaches provides a basis for
checks that can increase the trustworthiness of an implementation.

\section{Conclusion and discussion}
\label{SecConcl}
\subsection{Using optimisation and ODEs}
The determination of a profile likelihood confidence interval can be formulated
as a constrained optimisation problem. Since high quality optimisation routines
are widely available, this provides a practical way to compute confidence
intervals for the parameters or for a smooth function of the parameter
vector. This is especially interesting for models defined via formulas, as a
re-parametrisation is then generally difficult to achieve. A practical
difficulty is to ensure the convergence of the constrained optimisation, which
requires some fine tuning of the algorithm as well as suitable initial
values. This motivated the implementation of dedicated optimisation algorithms
by \cite{FisherLewis_Profile} and by \cite{BorisovMetelkin_Profile}. If MCMC
iterates are available corresponding to a reasonably non-informative prior, the
constrained optimisation formulation leads to a very simple determination of the
profile likelihood intervals, which can be useful when the log-likelihood
function is difficult to optimise. Wiping out the effect of the prior is then
easily achieved.

In the case where the smooth function depends on an extra continuous
``time'' variable, an ODE formulation based on the Lagrangian can lead
to some interesting theoretical and geometrical findings and provides
as well a solution to efficiently compute a batch of confidence
intervals, as often required in practice. Only one optimisation is
then required, in order to get an initial value.  The same technique
can be used to compute profile likelihood contours when the interest
is on a couple of parameters. Finally, by considering the confidence
level as being the time variable, we can derive as well an ODE that
exempts from the use of any constrained optimisation routine and
provides a very fast determination of the confidence interval(s). ODE
solvers are fast and require little tuning, hence may provide a
valuable alternative to constrained optimisation routines.

Interestingly, whatever be the parameter of interest, the ODEs used here have
their state vector being the parameter vector $\bs{\theta}$, possibly augmented
by a scalar Lagrange multiplier. While MCMC sampling uses a \textit{stochastic}
differential or finite-difference equation to explore the region of high
posterior density in the parameter space, we used here ordinary differential
equations to explore the region with high likelihood. These ODEs can be said
\textit{interest-driven} inasmuch they put emphasis on the parameter of
interest.

These techniques have been illustrated for models implying the GEV distribution,
with a small number of parameters. However, they apply to any parametric
model. A practical limitation is that the evaluation of the log-likelihood and
its derivatives must be fast enough to afford hundreds of evaluations, but this
limitation also holds for the inference based on MCMC.  Some further
experimentation will be required to better assess the practical acceptability
and the trustworthiness of the techniques described above. The ODE approach
could be compared to other methods or implementations and some porting between
the languages \Prolang{R}, \Prolang{Python} and \Prolang{Julia} could benefit to
a larger community of users.

\subsection{Limitations and possible extensions}

A practical limitation met both in the optimisation and the ODE method
is the need of scaling the parameters. However at least for EV
regression models, simple solutions could be used to avoid a manual
scaling. The QR decomposition or the singular value decomposition of
the design matrices such as $\m{X}^\mu$ both could be used.

As a theoretical limitation, no simple condition exists to grant that
a given function of the parameter can indeed be considered as a
parameter of the model as was assumed in the derivation. This relates
to the question of practical identifiability discussed in
\cite{FisherLewis_Profile} and \cite{BorisovMetelkin_Profile}. Only
sufficient conditions can be derived for some specific forms of
models, and especially for EV regressions with linear links as often
used in practice. An interesting property of the GEV distribution as
above parameterised is that the dependence on each parameter is
increasing for the stochastic ordering. For instance by
increasing~$\mu$ for given~$\sigma$ and~$\xi$, we get a distribution
which is larger in the stochastic ordering. So if by increasing a
given parameter~$\theta_k$ we are certain to increase~$\mu$, then we
will then also increase the distribution in the stochastic
ordering. This may be of some help in the analysis.

The ideas discussed here could be applied when the quantity of
interest depends smoothly on several variables $t_i$ e.g., takes the
form $\eta(t_1, t_2;\,\bs{\theta})$ for two variables $t_1$ and
$t_2$. Each bound of the confidence interval could then be obtained by
solving a vector-valued Partial Differential Equation (PDE) rather
than an ODE. The PDE can be derived by differentiating the KKT
conditions w.r.t. the variables~$t_i$.

The techniques described above could be used for Bayesian inference by simply
replacing the log-likelihood by the log-posterior, provided that the prior is
smooth enough and that its first and second order derivatives are
available. This may be useful e.g., if a high a posteriori probability contour
is to be found within a very small computing time. However, in order to control
the posterior probability level, some quadrature will be required.


In order to improve the coverage rate of the intervals in the small sample case,
a \textit{modified profile likelihood} can be used in place of the profile
likelihood, see \cite{Davison_StatisticalModels},
\cite{BarndorffnielsenCox_InferenceAsymptotics}. The confidence bounds related
to the modified profile can no longer be obtained by the constrained
optimisation described above, but the question arises whether similar
improvements could be reached by replacing the log-likelihood by a suitably
modified log-likelihood in the constraint.

\section*{Acknowledgement}

This work was partially funded by the French Institute for Radiological
Protection and Nuclear Safety (IRSN).  We are grateful to the members of the
IRSN \textit{Behrig} team 
for testing our R codes implementing the optimisation-based method described
here.

\bibliography{ConfInts}

\pagebreak

\appendix

\section{Proof of proposition~\ref{theo:Optim}}
\label{Proof}

  We first show that $\psi^\star \leqslant \Ub{\psi}$. By
  definition of the profile log-likelihood, we have
  $$
  \Prof{\ell}(\psi^\star) 
  = \ell\{\psi^\star,\,\hat{\bs{\lambda}}(\psi^\star)\} \geqslant 
  \ell(\psi^\star,\,\bs{\lambda}^\star)
  \geqslant \ell_{\max} - \delta, 
  $$
  because $\psi^\star$ and $\bs{\lambda}^\star$ together form the solution
  $\bs{\theta}^\star$ of problem (\ref{eq:PbStar}) which is in the feasible
  set. Since $\ell_{\max} - \delta = \Prof{\ell}(\Ub{\psi})$
  and since the profile log-likelihood is assumed to be decreasing for
  $\psi \geqslant \hat{\psi}$, we see that $\psi^\star$ is located at the left
  of the upper bound $\Ub{\psi}$ i.e.
  $\psi^\star \leqslant \Ub{\psi}$.
  
  \medskip\par\noindent 
  To see that $ \Ub{\psi} \leqslant \psi^\star $, observe
  that $\Ub{\psi}$ takes the form $\psi(\bs{\theta})$ for
  some $\bs{\theta}$ with
  $\ell(\bs{\theta}) \geqslant \ell_{\max} - \delta$, namely for
  $\bs{\theta}$ taken equal to the vector $\Ub{\bs{\theta}}$
  with components $\Ub{\psi}$ and
  $\hat{\bs{\lambda}}(\Ub{\psi})$. Indeed, then
  $\ell(\bs{\theta}_{\texttt{U}}) = \ell_{\texttt{prof}}(\Ub{\psi}) = \ell_{\texttt{max}} - \delta$
  by~(\ref{eq:ellDelta}). 
  Since
  $\bs{\theta}_{\texttt{U}}$ is therefore in the feasible set of
  problem~(\ref{eq:PbStar}), we have
  $\psi(\bs{\theta}_{\texttt{U}}) \leqslant \psi^\star$ i.e.
  $ \Ub{\psi} \leqslant \psi^\star $ as claimed.

\end{document}